\documentclass[
reprint, aps, pre
]{revtex4-2}

\usepackage{graphicx}
\usepackage{dcolumn}
\usepackage{bm}

\begin{document}
\title{Multicellular self-organization in \textit{Escherichia coli}}
\author{Devina Puri}
\email{dpuri@wustl.edu}
 \affiliation{Department of Molecular Microbiology, Washington University School of Medicine, Saint Louis, MO}
\author{Kyle R. Allison}
\email{kyle.r.allison@emory.edu}
 \affiliation{
 Division of Infectious Diseases, Department of Medicine, Emory University School of Medicine, Atlanta, GA
}
\date{March 3, 2025}

\begin{abstract}
\textit{Escherichia coli} has long been a trusty companion, maintaining health in our guts and advancing biological knowledge in the laboratory. In light of recent findings, we discuss \textit{E. coli}’s multicellular self-organization and develop general ideas for multicellularity, including the necessity for multicellular dynamics and interpretation by dynamic graphs, which are applicable to unicellular and multicellular organisms. In this context, we next discuss the documented behaviors of \textit{E. coli} self-organization (rosette formation, multicellular extension, and attached dormancy) and two potential behaviors (internal communication and mating). Finally, by comparing the dynamic graphs for different communities, we develop principles relevant to the theory of multicellularity.
\end{abstract}

\maketitle
\tableofcontents

\section{Background}
The bacterium \textit{Escherichia coli} has regularly advanced biological disciplines \cite{RN1, RN3, RN4, RN2}, even for higher organisms. Recently, the best-studied unicellular organism was discovered to perform surprising multicellular behaviors \cite{RN7, RN5, RN6}. Drawing on the wealth of existing knowledge for \textit{E. coli}, we expound its multicellular self-organization, by which we mean a dynamic process where the interactions between cells in a community are determined by their actions, without external control or constraint. Beforehand, we discuss \textit{E. coli}’s multicellularity, and develop ideas that may be useful for both unicellular and multicellular organisms. Though connecting disparate fields, our focus here is \textit{E. coli} self-organization, and we have not addressed many tangential topics.

\subsection{\textit{E. coli} multicellularity}
\textit{E. coli}’s “multicellularity” (\textit{i.e.}, the property of being a multicellular organism) has been considered \cite{RN10, RN8, RN9}, but it is not a multicellular organism. This partly motivated efforts to more rigorously define “multicellularity” through the lens of evolution \cite{RN11}. Bacteria can live \textit{multicellular-ly} at times, and have evolved forms of bacterial multicellularity \cite{RN12, RN13}, but a basis for comparing their multicellular behaviors is lacking. We think multicellularity should be a matter of degree—a measurable property rather than a categorical one—and we would define it as the complete set of an organism’s multicellular behaviors, each of which can be verified and listed. From this definition, the relative multicellularity of an organism is determined by the difference between two sets: its multicellular behaviors and those of another organism (we previously introduced “relative multicellularity” \cite{RN7} but less clearly). Fully specifying an organism’s multicellularity may be difficult, but identifying behaviors that distinguish it from others is often easy. To illustrate, we can say \textit{E. coli} has less relative multicellularity than the worm \textit{Caenorhabditis elegans} whose development includes creating organs; the loss of a multicellular behavior by evolution would reduce an animal's relative multicellularity; the programming of a new multicellular behavior by synthetic biology would increase an organism’s relative multicellularity; \textit{etc}. This definition of multicellularity is therefore broadly inclusive and can be applied to organisms that have no multicellular behaviors at all and to those that perform the largest set of multicellular behaviors. Moreover, multicellularity is then reframed as an empirical concept which can then be studied by genetics, developmental biology, and evolutionary biology, without having different definitions within each discipline. This framework is also useful for comparative multicellularity (analyzing behaviors across organisms) which is necessary to uncover fundamental principles, if they exist.

\subsection{Well-ordering of multicellular behaviors}
An obvious question follows from this way of thinking: are the behaviors of multicellularity well-ordered? That is—from the perspectives of evolutionary, synthetic, or developmental biology—do some behaviors always precede others? Such questions have been explored in cyanobacteria \cite{RN14} and solving them in a general sense will take time and collective efforts across disciplines. We believe such questions will be answered in the next decade and will reveal that much of multicellularity is well-ordered and that underlying rules govern relative changes in multicellularity. A common example is that complex multicellularity, associated with animals, only occurs in organisms that perform clonal multicellular self-organization \cite{RN15, RN11}, indicating this behavior is pre-requisite for complex multicellularity \cite{RN16}. Unexpectedly, we discovered that \textit{E. coli} performs a variation of clonal multicellular self-organization, unique to date among bacteria \cite{RN7}. \textit{E. coli}, which is better-studied than any other bacterium and simpler than any animal, lends itself to the basics of multicellularity, and the order of its behaviors \cite{RN5} may illuminate key rules.

\subsection{Constrained-organization and self-assembly}
A variety of \textit{E. coli} multicellular communities have been studied, though their differences and cell behaviors have not always been explicit or apparent. Pioneering studies investigated \textit{E. coli} colony formation on solid media at cellular resolution by light and electron microscopy \cite{RN17, RN18}, prompting thoughts about bacteria as multicellular organisms \cite{RN9}. Imaging techniques improved and the genetics and architecture of colony formation were investigated in greater depth \cite{RN19, RN21, RN20}.  Synthetic biology approaches have also used \textit{E. coli} colonies to program new multicellular behaviors \cite{RN22, RN24, RN25, RN23}, mimicking and modeling higher organisms. Being surface-bound, colonies are often described as biofilms and have been covered in related reviews \cite{RN27, RN28, RN26, RN10, RN9}. However, biofilms also tend to grow on surfaces in contact with liquid media. Experiments with fluid flow have been common in this area as it was reasoned this would increase collision of cells with surfaces and enhance biofilm formation \cite{RN29}. Such fluid flow biofilms are sometimes regarded as selection experiments \cite{RN30} as they are only formed by strongly-attaching cells that may be rare within populations. Commensal \textit{E. coli} does not form biofilms under fluid flow conditions \cite{RN31}, and there is a long history of studying \textit{E. coli} and \textit{Salmonella enterica} biofilms in static conditions, including pellicles which form at the air liquid interface after initiating at surfaces \cite{RN37, RN36, RN34, RN33, RN32, RN35}. Introducing conjugative plasmids that express adhesive pili can cause commensal \textit{E. coli} to form fluid-flow biofilms, but independently of type-1 fimbriae, curli, and other genes integral to wild-type formation of biofilms \cite{RN39, RN38}. \textit{E. coli} colonies and fluid flow biofilms, by definition and experimental practice, are usually clonal communities (\textit{i.e.}, wherein all members have derived from a single parental cell). Though they may differ in terms of their formation and genetics, they are also both constrained by their physical attachment to surfaces, an external feature that determines cell arrangement of cells in a community. Hence colonies, and related communities, result from constrained-organization.

There are also many examples of \textit{E. coli} auto-aggregation where microscopic communities, or “aggregates,” form without being constrained by surface contact. Examples of \textit{E. coli} aggregation typically result from introducing mutant adhesins (\textit{e.g.}, fimbriae \cite{RN40}),  engineered ones \cite{RN41}, or overexpressing native ones (\textit{e.g.}, \textit{Ag43} \cite{RN42, RN43}). By comparison, wild-type \textit{Pseudomonas aeruginosa} strains can aggregate without attaching to surfaces \cite{RN44}. Adding extracellular polymers enhances such aggregation \cite{RN45}, dependent on specific cell-surface molecules \cite{RN46}. As aggregate organization depends on the physical surface properties of cells and the density-dependent statistics of cell-cell collision, they have effectively been modeled as interacting particles by thermodynamics \cite{RN45}, equivalently to nanoparticle self-assembly \cite{RN48, RN47}. In these aggregates, the arrangement of cells is determined by their surface properties, rather than their actions, and on random collision events between separate cells. Such aggregates result from self-assembly rather than self-organization. Similar distinctions have been made in the past: self-assembly (processes tending towards equilibria, without requiring cell growth and adaptation) versus self-organization (non-equilibrium processes requiring cell growth and adaptation) \cite{RN49}.

The relative importance of different communities and their corresponding experimental methods is debated among those studying the multicellular behaviors of bacteria. This is to be expected when evidence to settle such debates is lacking, and until new data clarifies existing ideas, multiple conflicting interpretations all remain reasonable. The particular relevance of \textit{E. coli} self-organization as we have documented and discuss here is not yet fully clear (several possibilities are mentioned in \cite{RN5}), but it radically differs from past examples of bacterial multicellular organization.

\subsection{Self-organization and multicellular dynamics}
We recently discovered that \textit{E. coli} performs clonal self-organization of multicellular communities (see \cite{RN7, RN5, RN6}). This was unexpected: we had focused on tracking individual microbes, hoping to understand how antibiotic tolerance arose in communities like biofilms. Influenced by the old idea that biofilms might be “microbial development” \cite{RN50, RN51}, we decided to “developmentally” study \textit{E. coli} biofilms, that is to track the sequential events creating a multicellular community or organism from the first individual cell. From an engineering perspective, studying something developmentally is simply a question of correct scale and sampling. \textit{E. coli} cells are small ($\sim$2 $\mu$m long) and we therefore imaged at maximal optical resolution; \textit{E. coli} cells divide in $\sim$24 minutes, so with the Nyquist-Shannon sampling theorem in mind (\textit{i.e.}, reconstructing a digital periodic signal requires it to be sampled at least twice the rate of its frequency), we imaged every 6 minutes. We had anticipated observing constrained-organization, where individual cells attached to surfaces and grew by division into a multicellular colonies. Instead, we saw cells clonally organize into three-dimensional chain-like communities only to attach to surfaces and stop dividing once they contained $\sim$1,000 cells \cite{RN5}. We went on to find that the key genes that organize \textit{E. coli} biofilms \cite{RN34, RN35, RN52} all play unique stage-specific roles during self-organization, which itself was found to be responsible for producing biofilms. This process began with the formation of 4-cell rosettes, where a pair of cells stacked atop another and all of their long axes were parallel aligned.  How this organization occurred was unclear from the 6-minute data (another sampling problem), so we imaged every 2 seconds and discovered that  \textit{E. coli} cells use their flagellum to reposition relative to one another and thereby self-organize rosettes \cite{RN7}. We made these findings using experiments that allowed cells freedom of motion while also keeping them within the limited depth-of-field of high-magnification. Simply increasing the height and width of channels in the popular “mother machine” devices \cite{RN53} would allow similar tracking of bacterial self-organization, and multicellular behaviors more generally. Such systems could be used to study natural bacterial multicellularity or to troubleshoot multicellular synthetic biology approaches. In the future, similar behaviors will likely be identified in other bacteria once they are studied developmentally, by matching the rate of imaging to the dynamics of their cell behaviors. As studying bacteria “developmentally” will often reveal that their behaviors are not “developmental,” we prefer to call to this approach “multicellular dynamics” \cite{RN54}: the study of cell behaviors and changing cell-cell interactions in communities. Multicellular dynamics combines direct observations of cells in communities with genetics, molecular biology, and mathematical modeling, and is essential for accurate characterization of multicellular behaviors. To contextualize cell-cell interactions of multicellular dynamics, we think it is also useful to introduce dynamic graphs to represent multicellular communities.

\subsection{Dynamic graphs for multicellular communities}
Graph theory (often used interchangeably with “network theory”) is the study of sets of objects and the connections between them. It has been applied to biology, most commonly at the scale of objects within cells, \textit{e.g.}, metabolites, proteins, genes, and their interactions \cite{RN59, RN55, RN60, RN61, RN56, RN58, RN57}. Generally, the individual nodes of the graphs (\textit{e.g.}, proteins) and their connections (\textit{e.g.} protein-protein interactions) are fixed and in a sense encoded by DNA. Such graphs are static, their nodes and connections do not change with time. In multicellular communities however, new nodes (cells) are continually produced and their connections (cell-cell interactions) naturally change with time as well. The communities and cell-cell interactions of multicellular dynamics must therefore be considered as dynamic graphs \cite{RN62} (where nodes and vertices are added or removed over time).

Dynamic graphs require propagation rules for the addition or subtraction of nodes and connections \cite{RN62}. For most multicellular communities, two nodes are created from an existing node by cell division, and their connections are determined by their position after division. Connections can represent physical contact between cells by their appendages or communication by diffusible molecules. Connections between neighboring cells are more impactful than those between distantly positioned cells and the connectivity of multicellular graphs is sparse compared to others used in biology, like gene networks \cite{RN63, RN64}. Throughout the figures here, we introduce dynamic graphs for multicellular self-organization, as the roles of appendages in organizing \textit{E. coli} communities previously represented \cite{RN7, RN5, RN6}. Even without theoretical development, which will be straightforward in cases, dynamic graphs suggest novel insights into multicellularity. Promisingly, dynamic graphs have been applied to both community fracturing in yeast \textit{Saccharomyces cerevisiae} \cite{RN66, RN65} and morphogenesis of plant hypocotyl \cite{RN67}, and more broadly recommended for developmental biology and multicellular evolution \cite{RN68}. Separately, their application to neural networks is state-of-the-art in machine learning \cite{RN69}.

A defining aspect of biology is that cells sense their environment and respond by altering their behavior and chemical composition. In communities, cells can sense their interactions and respond by changing them, thereby re-writing the rules of their multicellular organization. For multicellular graphs, this means the rules of propagation from one graph to the next are also dynamic and depend on community stage. This has methodological implications. Though they are commonly sought and applied, “general” rules which are static in time have limited ability to explain multicellular self-organization, which is necessarily adaptive and changing. The propagation rules for multicellular self-organization are instead “particular,” they change with time and from organism-to-organism. Theoretically deriving such rules will often be unreliable, and they must instead be empirically discovered, by tracking cells and their changing interactions in communities (\textit{i.e.} by multicellular dynamics). Without direct verification of cell behaviors, myriad theoretical postulates cannot be distinguished. This point may also suggest limits on the ability of artificial intelligence (AI), which extrapolates general patterns from existing data, to understand multicellular biology. AI cannot use particulars that have not been observed, though we concede it may already be better than a review article at summarizing current knowledge in microbiology.  Moving on, we next cover the documented stages of \textit{E. coli} self-organization.

\section{Documented behaviors in \textit{E. coli} self-organization}
\subsection{Rosette self-organization}
Single cells form rosettes by a choreographed dynamical process involving cell division, cell-cell adhesion and cell repositioning \cite{RN7, RN5} (\textbf{Fig. 1}, left).
\begin{figure}
\includegraphics{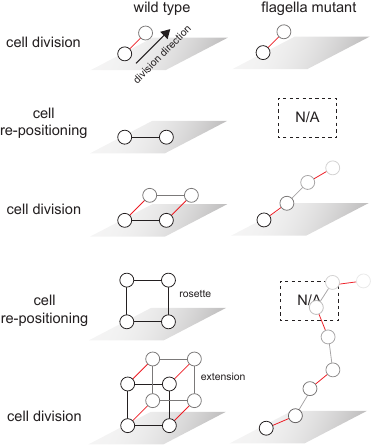}
\caption{Rosette self-organization.
\textbf{(Left)} Graphs illustrating the multicellular dynamics of rosette formation by \textit{E. coli}. Nodes indicate cells and vertices indicate cell-cell connections. \textit{E. coli} is rod shaped and the direction of cell division is determined by the orientation of cells. New cell-cell connections created by cell division are indicated in red. These new connections are mediated by\textit{Ag43} (encoded by \textit{flu} gene) at the poles of sister cells. Re-positioning pivots sister cells about their point attachment to align with their long axes parallel. This motion follows a random walk through angular space and is produced by the flagellum. The “rosette” stage and first step of “multicellular extension” to tube-like chains are indicated. The square-like configuration achieved by rosettes is approximately maintained throughout extension (\textbf{Fig. 2}) and attachment (\textbf{Fig. 3}) and further cell re-positioning is not generally observed (caveat in \textbf{Fig. 2}). Rosettes are clonal are not attached to surfaces. Multi-clonal communities can however occur in very high densities if separate cells aggregate. \textbf{(Right)} Graphs illustrating the multicellular dynamics of organization by flagella mutants (either $\Delta$\textit{fliC} or $\Delta$\textit{motA}) \textit{E. coli}. These cells do not reposition and communities grow in single-cell thick chains where cells adhere at their poles. The angle between cells in flagella mutants is evidently random, and cell folding can technically occur but is statistically rare.}
\end{figure}
Adhesion occurs at the newly-formed poles of sister cells after division. It is mediated by Antigen 43 (\textit{Ag43}), a self-recognizing surface protein that binds to its cognates on closely neighboring cells \cite{RN42, RN72, RN70, RN71}. Though \textit{Ag43} can be located anywhere in the membrane, it migrates to the cell poles for sister cell adhesin \cite{RN73}. Septation, the invagination of the cell envelope as one cell splits into two during division, may facilitate sister cell adhesion by bringing \textit{Ag43} molecules into the correct orientation to bind to one another. \textit{Ag43}, encoded by the \textit{flu} gene, consists of two main domains: an N-terminal $\alpha$ subunit responsible for adhesive properties and a C-terminal $\beta$-barrel domain that anchors the protein to the outer membrane \cite{RN72, RN70}. The $\alpha$-subunit reaches the cell exterior, while remaining attached to the $\beta$ subunit at the other end. The adhesive interactions of \textit{Ag43} are considered a Velcro-like mechanism, where the binding of \textit{Ag43} proteins on adjacent cells pulls the cells together draws cells together further increasing adhesive interactions \cite{RN74}. It plays a crucial role in cell adhesion and aggregation, promoting multicellular clusters and biofilms \cite{RN37, RN71, RN75}. \textit{Ag43} provides very short range interactions, $\sim$10 nm \cite{RN74}, which are blocked by other extracellular appendages \cite{RN76, RN43, RN73}. K-12 \textit{E. coli} is missing larger surface antigens common in wild strains \cite{RN79, RN78, RN77} so this may not often present an issue. However, wild strains, like uropathogenic \textit{E. coli} (UPEC) have an array of antigens and adhesins that could present a barrier to \textit{Ag43 }interactions. But rather than dispensing with \textit{Ag43}, UPEC has two versions of it with different binding affinities and both play critical but distinct roles in pathogenesis \cite{RN74, RN75}. The cell envelope at the poles is formed rapidly by septation, and initially there would be no appendages to maintain adherence between sister cells. This also means that there would be little to block \textit{Ag43} interactions at the poles of new sister cells, which may then be a specialized function that enables multicellular communities to form in a clonal manner. \textit{Ag43} is important for sister-cell adhesion and rosette formation, and $\Delta$\textit{flu} strains do not form clonal chains. At high cell density however, $\Delta$\textit{flu} cells can aggregate to form polyclonal communities that approximate some aspects of multicellular extension.

At the 2-cell stage, after dividing and adhering, sister cells reposition relative to each other using their flagellum \cite{RN7}. Sister cells are relatively motionless for up to $\sim$10 minutes after division, before rapidly folding onto one another. A similar folding of sister cells after division was previously reported \cite{RN80}, but before the advent of automated imaging in microscopy. The orientation of the cells at this time can be thought of as two vectors, each $\sim$2 $\mu$m long, with tails joined and their respective arrow heads pointing in opposite directions. The angle between these cell vectors is initially 180 degrees. During folding, one of the cells performs an angular random walk to cover the entire 180-degree angle and lay on top of the other sister cell. The cell vectors stack parallel and the angle between them becomes 0 degrees. The folding motion is comprised of directional strokes, where a cell pole travels approximately 1 $\mu$m at a time, interspersed within pauses during which the direction of the next stroke is randomized. Flagella are helical filaments rotated by a reversible motor, to swim and produce motility in bacteria. The motor of the flagellum is ionic and allows it to rotate at high speeds, $\sim$300 rotations per seconds, which in turn propel the cell forward. The flagellum has three main components: the basal body, hook, and filament \cite{RN81, RN82}. The basal body includes rings embedded in the cell membrane and contains the motor, the hook serves as a connector, and the filament, composed of flagellin, is the propelling helical tail that extends from the cell surface. When rotating counterclockwise, synchronously as a bundle, \textit{E. coli} cells move in a linear swimming direction. On the other hand, switching to clockwise rotation unravels the bundle, causing the cell to tumble.

 After folding, cell pairs synchronously divide with the newly created sister cells adhering to one another at their poles. This is the logical result of the first folding event as it aligns the division planes of two cells in parallel and their division cycles will be closely synchronized as they just divided form the first initial cell. Often immediately after division (rather than at a 10 minute delay), these communities perform a second folding event, again about their point of contact at the cell poles, to align all four cells in parallel is a square or quatrefoil like configuration. Intuitively and from visual inspection, the motion of the next folding motion has only one degree of freedom whereas the first had two. Cell pairs move back and forth, as would the hinge of a door or the wings of butterfly. Both folding events, once motion begins, generally occur in under one minute but can take place in a matter of seconds. The second folding even, at the 4-cell stage, established quatrefoil-configured rosettes, and there is little cell repositioning noted afterward during self-organization. The process of rosette formation is robust in hydrostatic environments and was observed in the majority of cells tracked by microscopy \cite{RN7}.

 Precise cell propulsion and angular movements generated by the flagellar motor are critical for rosette self-organization. In the absence of flagellar activity (by knocking out genes for the filament or the motor), reliable repositioning of sister cells is perturbed, preventing the establishment of rosette geometries \cite{RN7} (\textbf{Fig. 1}, right). Instead, these communities form long chains where cells are adhered at their poles and do not precisely control their interactions. Conversely, excessive cell motion disrupts polar adhesion between sister cells, leading to their separation as individual cells. The cell repositioning of rosette self-organization balances cell adhesion and propulsion, and it is interesting to note recent findings that flagellar motility is modulated, but not abolished, in an adherent-invasive \textit{E. coli} strain associated with Crohn’s disease \cite{RN83}.

 The cell-cell configuration of rosettes creates a tube-like inner cavity with a diameter of $\sim$1-2 $\mu$m \cite{RN5} and a volume of $\sim$1-3 femtoliters ($\mu$m\textsuperscript{3}), open on both ends. Multicellular extension approximately maintains this configuration, the width of communities, and hence the diameter of the inner cavity. The adhesins involved in extension and attachment (type-1 fimbriae, curli, and polyglucosamine) do not play a role in rosette self-organization, as strains with mutations to them form rosettes identically to wild-type strains. This could be by design: structurally rigid fimbriae, which grow to 1-2 µm, would sterically block cell repositioning if they were present; and sticky extracellular matrix components (curli and polyglucosamine) would reduce cells’ freedom of motion and might additionally attach them to surfaces or adhere them to other cells. Such variations would alter the multicellular organization of communities in ways that may be predictable, which may be relevant to communities in other species, many of which share similar sets of extracellular appendages. Quickly creating internal cavities is a hallmark of development in multicellular organisms: it aides cell communication, coordinated behaviors, differentiation, and nutrient distribution \cite{RN84}. The efficiency of rosette formation during \textit{E. coli} self-organization may reflect a primitive variant of these advanced multicellular traits. Though primitive, it is not a precursor as \textit{E. coli}, like many other microbes, evolved in a world already dominated by multicellular organisms. To live amongst such organisms, microbes may at times have mimicked some of their behaviors. Studies support the essentiality of rosettes to embryogenesis, organogenesis, and neuronal development \cite{RN87, RN88, RN89, RN86, RN85}. Rosette formation has been studied in simple organisms as well, like the unicellular choanoflagellates \cite{RN90} which are relatives of the ancestors of early animals. Rosettes have also been described in other bacterial species \cite{RN97, RN98, RN94, RN91, RN93, RN95, RN96, RN92}, but there remain many questions regarding their formation and consequences, which will likely be as diverse as the species creating them. Studying the multicellular dynamics of bacteria, tracking the behavior of cells at the appropriate resolution, will elucidate the significance of rosettes, to infections, antibiotic resistance, and other phenomena. 

\subsection{Extension and reproduction}
Extension begins from rosettes which are $\sim$2-$\mu$m long and contain 4 cells until communities are $\sim$200-400-$\mu$m long and contain up $\sim$1,000 cells \cite{RN5} (\textbf{Fig. 2}).
\begin{figure}
\includegraphics{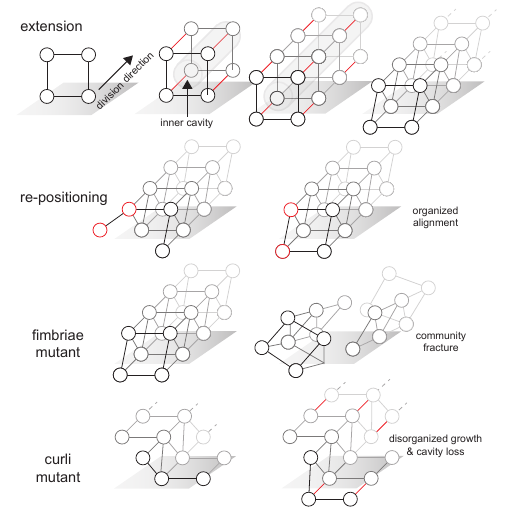}
\caption{Multicellular extension.
\textbf{(Top)} Graphs illustrating rosette extension into tube-like multicellular chains by \textit{E. coli}. Nodes indicate cells and vertices indicate cell-cell connections, red depicting new ones created by cell division. Multicellular chains propagate exponentially in length while their width remains constant, thereby maintaining the diameter of the inner cavity (depicted) between cells. This is a result of the division direction of all cells being aligned within rosettes and their cell cycles being closely synchronized. Fimbriae coordinate stable cell-cell interactions and curli coordinate cell-cell organization necessary to maintain the inner cavity. Cells can marginally shift within chains, increasing the total cell-cell connections (far right), but they remain aligned in parallel and communities maintain their square-like configuration. A unique property of propagating by extension is that each cell has access to both the inner cavity and the extracellular environment, regardless of the number of cells. \textbf{(Upper middle) }Graphs illustrating cell repositioning during extension. Occasionally new cells (depicted in red) are not immediately adhered to their neighbors. These cells can re-position by the same movements that form rosettes, causing new cells to adhere to their neighbors in the correct alignment. Cell repositioning during extension has only been observed during its early phase (by 32 cells). \textbf{(Lower middle) }Graphs illustrating the multicellular dynamics of fimbriae mutant ($\Delta$\textit{fimH}). These cells form rosettes and begin extension, but cell-cell interactions are unstable and communities fracture into smaller multicellular clusters. If the density is high enough, these clusters can aggregate and recapitulate aspects of wild-type extension and attachment. \textbf{(Bottom) }Graphs illustrating the multicellular dynamics of curli mutant ($\Delta$\textit{csgA}). These cells stably adhere to each other but lose control of their organization: they form rosettes and begin extension but communities lose rosette configuration and flatten, thereby also losing their inner cavity. Ultimately, these communities produce little to no attachment to surfaces.}
\end{figure}
Throughout this phase, the average width of multicellular communities, which resemble freely moving chains, is constant at $\sim$3 $\mu$m (diameter may increase by $\sim$0.5-1 $\mu$m, but cannot be statistically proven from the existing data). As the relative configuration of cells is approximately maintained, extension by default also extends the inner cavity enclosed within communities (\textbf{Fig. 2}, top). Though the inner cavity is extended, rosette configuration is not perfectly maintained as the poles of cells stacked on top of each other shift during extension, which may be due variations in cell growth or division dynamics (\textbf{Fig. 2}, top right). Generally, cells do not perform repositioning relative to their neighbors in multicellular chains. The caveat is that cells at the ends of growing chain are sometimes out of alignment with their neighbor cells. In this case, such cells perform a folding behavior similar to that observed during rosette formation that brings them into alignment with the neighboring cells (\textbf{Fig. 2}, middle top). Hence, though mostly lacking after rosette formation, flagella-mediating folding appears to self-regulate the multicellular geometry of communities during extension.

In addition to maintaining an inner cavity, extension also maintains exposure of each cell in the community to the external environment. This may be critical for acquiring nutrients and growing. Relatedly, extending communities grow at an exponential rate, approximately equivalent to that of single cells in the same growth medium \cite{RN5}. In bacteria, the formation of multicellular communities is often thought to require sacrificing rapid growth \cite{RN99}. This is not the case for \textit{E. coli} self-organization. During extension, chains can fracture into two or more smaller multicellular communities. Community fracturing has been considered a basic form of reproduction in examples of simple multicellularity \cite{RN65, RN100}. This appears to apply to \textit{E. coli} self-organization also: fractured chains generally extend as their parental chains had and form new independent chains. Fracturing therefore enables self-propagating. As extension and growth have a clear end point during self-organization when chains attach to surfaces and cease dividing, the creation of new chains might ensure that some members of a community continue growing. Therefore, it could represent an ideal example of bet-hedging, and previous theoretical development \cite{RN101} suggests fracturing rate and extension duration would be determined by the frequency of changes in \textit{E. coli}’s environmental niche. In this case fracturing rate depends on the strength of cell-cell adhesion in communities and external or cell-produced forces. This self-propagation of multicellular chains appears to also propagate biofilms, based on comparing the robustness of biofilm formation in different strains to varying initial cell densities \cite{RN5}.

Chains however do have moderate stability during extension as they remain clonal and do not appear to adhere to free-floating cells or those from other chains. Clonality may offer several benefits. It ensures all cells are genetically identical, thereby reducing genetic conflict and promoting cellular cooperation \cite{RN103, RN102}. This uniformity ensures that cells can predictably coordinate their behaviors to develop multicellular traits \cite{RN103, RN102}. From the perspective of evolutionary biology, clonality shifts the focus of selective pressures from the level of individual cells to that of groups \cite{RN103, RN102}, which is thought to be essential to evolve traits that improve the fitness of the entire organism. Clonality is also critical for the predictable and ordered propagation of multicellular communities (discussed below in Comparative multicellularity). Overall, clonality helps communities coordinate the behaviors of their individual cells and is the rule for complex multicellular organisms. \textit{E. coli} is not that complex, but its preference for organizing clonal communities suggests that these benefits may be general for multicellularity and the behaviors comprising it. Methodologically, clonality had other benefits. It was the rationale for developing the rather simple method for revealing cellular arrangement in biofilms by mixing identical cells harbored either red or green fluorescent proteins \cite{RN6}, and thereby connecting \textit{E. coli}’s multicellular dynamics to the organization of bulk scale biofilms. Though experimentally straightforward, biofilms had not been investigated by using the same strain but with two colors, though similar approaches are common in studying range expansion within colonies on solid media \cite{RN106, RN104, RN105, RN107}.

Genetically, extension relies on at least two adhesins (type-1 fimbriae and curli) in addition to \textit{Ag43}, which likely continues to facilitate polar attachment of new sister cells. Fimbriae are rigid hair-like appendages that extend radially from the surface of many gram-negative bacteria, including \textit{E. coli} \cite{RN109, RN108}\textit{.} Their essential role in hydrostatic “pellicle” biofilms in gram-negative bacteria has long been known and they are regulated in a switch-like manner induced by growth in static cultures \cite{RN111, RN112, RN33, RN32, RN110, RN113}. Fimbriae play a crucial role in bacterial adhesion, colonization, and pathogenesis \cite{RN108, RN114}. Typically, fimbriae are 7-10 nm in diameter and $\sim$1-2 $\mu$m long, have a helical structure, and often contain adhesins at the tip \cite{RN40, RN115}. These adhesins, such as \textit{FimH} in type-1 fimbriae, can bind to specific receptors (\textit{e.g.}, mannose) on host cells, facilitating invasion of tissues \cite{RN40}. The assembly of fimbriae involves the chaperone-usher pathway, where pilin subunits are transported to the periplasm and stabilized by chaperone proteins \cite{RN114}. The usher protein in the outer membrane facilitates pilus assembly and extrusion \cite{RN114}. Fimbriae are critical virulence factors in uropathogenic \textit{E. coli} and facilitate the formation of intracellular bacterial communities (IBCs) within bladder epithelial cells, protecting bacteria from immune responses and antibiotics \cite{RN114}. Fimbriae are also essential for \textit{E. coli} biofilm formation \cite{RN34, RN35} without the presence of host cell receptors, indicating they also facilitate cell-cell adhesion in \textit{E. coli} communities.

During extension, fimbriae help stabilize chains \cite{RN5} (\textbf{Fig. 2}). \textit{FimH} mutants ($\Delta$\textit{fimH}), lacking the adhesive component of fimbriae, fracture during extension, which in some chains can occur as early as the 16-cell stage. The length of fimbriae is approximately equivalent to the diameter of the inner cavity during extension and lateral interactions between neighboring cells by fimbriae would necessarily restrict relative cell movement during extension. Practically, if growing to their full lengths, fimbriae would structurally expanded the inner cavity and ensure that it did not collapse. Moreover, $\Delta$\textit{fimH} strains are thought to still make rigid fimbriae but their adhesiveness is blocked \cite{RN37}. From this perspective, the fracturing of $\Delta$\textit{fimH} strains may result from fimbriae pushing cells apart, blocking \textit{Ag43 }adhesion without replacing it with \textit{FimH} adhesion. 
 Similar to $\Delta$\textit{flu} cells, $\Delta$\textit{fimH} cells can form polyclonal aggregates when cell density is high and approximate aspects of self-organization. Future investigation by electron microscopy are needed to clarify the function of fimbriae in self-organization.

 Curli are proteinaceous extracellular fibers produced by many \textit{Enterobacteriaceae}, including \textit{E. coli} \cite{RN118, RN117}. They are essential components of biofilm extracellular matrix, and they aid in adhesion and biofilm formation, host cell invasion, and colonization and persistence in host tissue \cite{RN121, RN118, RN122, RN120, RN119, RN34, RN117, RN21}. Composed primarily of \textit{CsgA} protein subunits and nucleated by \textit{CsgB}, curli fibers are assembled with the help of accessory proteins encoded by the \textit{csgBA} and \textit{csgDEFG} operons \cite{RN121}. Curli assembly starts with \textit{CsgA} and \textit{CsgB} secretion via \textit{CsgG} and facilitated by \textit{CsgE} and \textit{CsgF}. Regulation of curli gene expression involves environmental cues and the transcriptional regulator \textit{CsgD} \cite{RN121}. Differing from fimbriae, curli polymerization take place extracellularly at a rate dependent on the external concentration of its subunits. Hence, it should be densest near the cell surface, evident the ECM cocoons encasing cells in biofilms \cite{RN120}. Similarly it would rapidly polymerize in the inner cavity during extension, and it would affect this spaces shape and the diffusion properties. During extension, \textit{csgA} mutants do not fracture as \textit{Ag43} or fimbriae mutants, and their communities are generally clonal. However, they do lose the initial configuration between cells established by rosette formation and communities fold open and lose their inner cavity \cite{RN5} (\textbf{Fig. 2}). These communities, though their physical interactions between cells would appear stable, dissociate in small cell clusters rather than attach to surfaces at the end of extension.

Extension may be a general behavior
. A wild type strain of \textit{Bacillus subtilis} makes similar multicellular chains (which we estimate are extended at near-constant width) during biofilm formation \cite{RN124, RN123} which in turn requires amyloid fibers (like curli) \cite{RN125}. Regulating and remodeling extracellular matrix in an internal cavity is essential for variety of developmental behaviors \cite{RN89}, as seen in choanoflagellates \cite{RN126, RN90} or to “open up” and advance the lumen during mouse embryogenesis \cite{RN127}.

\subsection{Attached dormancy}
Chains attach to surfaces once they extend to a length between 200-400 $\mu$m as tracked dynamically in devices and $\sim$400 $\mu$m when observed in biofilms, corresponding to $\sim$1,000 cells in communities \cite{RN5}. Attachment stabilizes cell-cell interactions in communities, especially relative to the moderate stability of chains during extension, during which they periodically fracture. Rinsing surfaces where biofilms are growing leaves full length ($\sim$400 $\mu$m) chains attached to surfaces but removes shorter chain-like communities 50-100 $\mu$m long, which are imputed to be extending chains that have lost their multicellular configuration by being compressed by cover glass. Biofilms communities are similarly compressed by cover glass but their attached chains do not lose their multicellular configurations, further indicated attachment stabilizes cell-cell interactions. Attachment coincides with dormancy: quantifying the motion and area of chains demonstrated that their movement and growth stop simultaneously \cite{RN5}. Considerable support also comes from biofilm experiments showing chains once attached and that biofilms themselves grow through successive attachment of chains on top of one another. Though no longer actively dividing, cells are not fully dormant: they can be killed by metabolite potentiation of aminoglycosides \cite{RN128} indicating they maintain a degree of both metabolism and protein synthesis. Though life goes on for these communities, attached dormancy is the end-point of their self-organization. It may also represent the end of \textit{E. coli}’s multicellular dynamics given that cells no longer divide and their behaviors and interactions may change little. Such properties are also characteristic of differentiated cells in multicellular organisms, which carry out specific functions though they do not divide or change much. Though self-organization’s end-point resembles a differentiated state, we prefer “attached dormancy” for now until evidence clarifies its function.

 Attachment results from production of poly-$\beta$-1,6-N-acetyl-D-glucosamine (PGA), a linear extracellular polysaccharide composed of unbranched $\beta$-1,6-linked N-acetyl-D-glucosamine (GlcNAc) residues \cite{RN129, RN52}. Genes for producing and exporting this polysaccharide are encoded by the \textit{pgaABCD} locus in \textit{E. coli }\cite{RN52}. During biofilm formation, PGA promotes attachment of bacterial cells (\textit{E. coli}, \textit{Pseudomonas fluorescens}, and \textit{Staphylococcus epidermidis}, but not \textit{P. aeruginosa} or \textit{Salmonella enterica}) to abiotic surfaces, facilitates intercellular adhesion, and makes up part of the extracellular matrix \cite{RN129, RN52}. PGA may be sufficient for attachment: a comprehensive study did not identify requirements for the other adhesins for surface attachment when PGA was supplied exogenously \cite{RN130}. This would indicate that when PGA is present, the other adhesins are dispensable for attachment, though the production of PGA during self-organization first requires other adhesins to properly function during rosette formation and multicellular extension.
\begin{figure}
\includegraphics{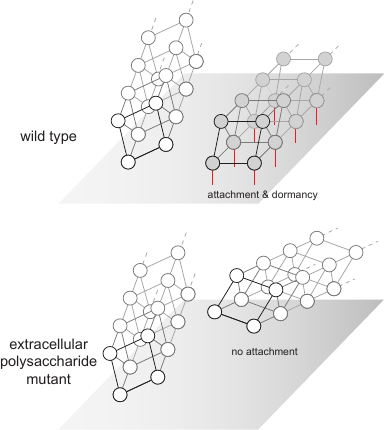}
\caption{Attached dormancy.
\textbf{(Top)} Graphs illustrating attachment of tube-like multicellular chains by \textit{E. coli}. Nodes indicate cells and vertices indicate cell-cell connections. Filled in nodes indicate cells that have halted division and red vertices indicate attachments between cells and surfaces. These attachments are mediated by the extracellular polysaccharide (poly-N-acetylglucosamine). Division of cells within communities halts at the same time as community attachment, when chains are $\sim$200-400 $\mu$m in length and have $\sim$1,000 cells. Prior to attachment, during rosette self-organization and extension, chains are freely moving and have little to no surface attachment. Repeated, parallel-aligned attachment of tube-like chain communities produces biofilms. \textbf{(Bottom) }Graphs illustrating the multicellular dynamics of extracellular polysaccharide mutant ($\Delta$\textit{pgaB}). These cells perform rosette self-organization and multicellular extension, but do not attach to surfaces and do not appear to halt cell division. Mutations interrupting rosette formation (\textbf{Fig. 1}) or extension into chains (\textbf{Fig. 2}) also impair surface attachment, and the effect of each mutation on multicellular self-organization is consistent with previous findings on \textit{E. coli} (K-12) biofilm formation. }
\end{figure}

In self-organization PGA production is required for chains to attach to surfaces (\textbf{Fig. 3}). Mutant strains for PGA production perform rosette formation and multicellular extension equivalently to wild-type, but simply fail to attach and stop moving at the 1,000 cell stage. Instead, communities flexibly move and they dissociate into individual cells and small clumps approximately an hour after missing their attachment time \cite{RN5}. PGA production has a significant metabolic cost \cite{RN131, RN52} which may partly explain why dormancy and attachment coincide. On the other hand, this cost is likely to be transient, and division may cease for other reasons. One possibility is that surface attachment might limit nutrient access for some cells. Though another possibility is suggested by the expression of the master regulator \textit{rpoS}, which regulates many dormancy related genes \cite{RN132, RN119}, $\sim$1-2 generations prior to attachment \cite{RN7}: perhaps dormancy is the end goal of self-organization and attachment is one part of a bigger program (see also Comparative multicellularity and dynamic graphs below). Mutations that impair rosette formation (\textit{e.g.} to \textit{Ag43} and flagella) or multicellular extension (\textit{e.g.} fimbriae and curli) and negatively impact surface attachment of communities that happen to have formed by aggregation. It is unclear if this is because each stage of self-organization requires the preceding stage (to properly regulate stage-specific gene expression) or if surface attachment requires the other adhesins to work in concert with PGA (this seems unlikely given the sufficiency consideration above). There is also a symmetry to \textit{E. coli} self-organization ending with attachment to a surface, as the entire process may presuppose a nearby surface. Rosette formation (with its balance of forces) and multicellular extension (with its moderate stability) require hydrostatic conditions, which are common in nature and generally encountered near physical surfaces. In a sense, \textit{E. coli}’s self-organization (from the first to the thousandth cell) requires a local surface, which it ultimately chooses to attach to rather than continue dividing. Why \textit{E. coli} does this is still unclear, though attaching to a rigid and stable surface is a good way for multicellular communities to fix their cell-cell interactions for the long-term. Future investigations will clarify some of these ideas and identify additional multicellular behaviors performed during self-organization.

\section{Potential behaviors in \textit{E. coli} self-organization}

Characterization of \textit{E. coli} self-organization has only just begun and as it proceeds, other multicellular behaviors will likely be revealed. Self-organization ends in surface-attached biofilm communities and explains the roles of key biofilm genes; we therefore suspect properties attributed to macroscopic biofilms, like antibiotic tolerance and toxin secretion, are in fact due to microscopic self-organization. Expression during self-organization of alternate sigma factor \textit{rpoS} \cite{RN7}, which regulates critical properties across diverse species \cite{RN132, RN119}, increases this likelihood. We discuss two other potential forms of cell-cell interactions: internal communication and mating.

\subsection{Internal communication}

Bacteria produce diffusible molecules that communicate between cells \cite{RN133} and coordinate their behaviors, sometimes separately classified as “signals” and “cues” \cite{RN135, RN134}. “Quorum sensing” molecules, like auto-inducer 2 (AI-2), are widely known examples \cite{RN136} and production of AI-2 by \textit{LuxS} in \textit{E. coli} was reported more than 25 years ago \cite{RN138, RN137}. Sensing of external AI-2 occurs via the genes of the \textit{lsr} operon \cite{RN139} and has been associated with drug resistance, phage defense, chemoreception, motility, osmotic stress, and others in commensal \textit{E. coli} \cite{RN140} and it also controls type-III secretion in enteropathogenic \textit{E. coli} \cite{RN141}. Despite the many associations, the precise function of AI-2 sensing remains unclear \cite{RN133}. That it is also associated with biofilm formation \cite{RN140} suggests it could be reported during self-organization in the future.

“Quorum sensing” as a term was proposed given that sufficiently large populations (“quorums”) of bacteria were required for auto inducer activity in well-mixed broth cultures of planktonic cells \cite{RN142}. Recently, microscopic \textit{E. coli} communities adhered by \textit{Ag43} were shown to sense AI-2, measured by a \textit{lsr} transcriptional reporter, while nearby planktonic cells in the same cultures did not \cite{RN143}. This means, in multicellular communities, \textit{E. coli} senses AI-2 before the overall population reaches the requisite “quorum.” Similar findings have been reported for \textit{P. aeruginosa} \cite{RN144}. AI-2 therefore can function as a multicellular signal in \textit{E. coli} communities, even small ones (from the data \cite{RN143}, we estimate communities contain $\sim$30 cells or at least between 4 and 100 cells). Communities this size contribute to infections \cite{RN54, RN145}, but are seldom studied in the lab. For molecules like AI-2, the “quorum” needed in liquid cultures \cite{RN142} could result from cells trying to sense a multicellular signal while being separated into individuals by external forces.

Multicellular communication by diffusible molecules is common in higher organisms, and its relevance to development was perhaps first famously elaborated by Alan Turing \cite{RN146}, who illustrated multicellular patterns could theoretically arise by tuning diffusion parameters. This idea predates Jacques Monod’s and Francois Jacob's work on gene regulation which would add to it by demonstrating how cells dynamically alter the molecules they produce \cite{RN147}. And an empirical approach to development later led Lewis Wolpert to suggest cells in communities have positional information (they know “who” and “where” they are) which they can alter and interpret, partly by communicating through diffusible molecules \cite{RN148, RN149}. Biological communication by diffusion has characteristic distances, which are short rather than long \cite{RN150}. The time for a molecule to diffuse between two cells increases with the square of the distance: \textit{i.e.}, a molecule takes 100 times longer to diffuse between cells when they are 10 times farther apart. Hence, controlling the distance between cells is fundamental to communication, which occurs at the microscopic scale. It typically takes place between direct neighbors who employ strategies to regulate their distance. For example, the synaptic cleft is maintained at $\sim$20 nm \cite{RN150} to maximize the speed and robustness of neuronal signaling. In development, the distance between cells can be regulated by creating inner cavities (rosettes, lumens, tubes, \textit{etc.}), which impose boundary conditions that concentrate diffusible molecules and create a space for internal communication.

But communities grow during development, often exponentially, presenting challenges for internal communication. In spherical communities, seen in some simple multicellular eukaryotes \cite{RN90, RN151}, the inner volume increases faster than the number of cells; a molecule’s internal concentration would plummet unless each cell increased its production. Such geometries necessarily limit internal communication. The obvious solution is to regulate the diameter of cavities as communities grow, effectively fixing the ratio between volume and the number of cells. Organizing as multicellular tubes that propagate lumens (cylindrical inner cavities) achieves this and is common in the development of higher organisms \cite{RN150}. By maintaining critical dimensions for diffusion, lumen extension therefore enables robust communication within growing multicellular communities. Production of the extracellular matrix, a interwoven mesh of proteins and polysaccharides, within lumens further controls communication; it can selectively limit diffusion of molecules and concentrate others near cell surfaces \cite{RN150}. Taken together, internal communication determines multicellular organization and also the form of organization fundamentally determines communication. Previously, it was suggested that rosettes may be a key unit of development \cite{RN87}. But without transitioning into lumens, rosettes could not maintain the robustness of communication and would lose control of their future development. Tubes, lumens, and other constant-diameter geometries, which rosettes happen to initiate, may alternatively be an essential strategy to development.

AI-2 signaling by the inner cavity has not been reported in \textit{E. coli} self-organization (\textbf{Fig. 4}, top). But it may be in the future: \textit{E. coli} actively stabilizes and extends its inner volume while adapting its properties by extracellular matrix components like curli \cite{RN5}. Additionally, \textit{E. coli} senses AI-2 in even small multicellular communities \cite{RN143}. The growth rate and boundaries conditions of \textit{E. coli} self-organization \cite{RN5} could be combined with theoretical approaches to quorum sensing \cite{RN152, RN153} to develop dynamic models for \textit{E. coli} multicellular communication. Such models would not replace experiments, but would help guide and interpret them. The selective pressures that drove \textit{E. coli} to perform these behaviors, and if they were intended for multicellular communication, are considerations for evolutionary biology. Though it is worth noting many bacterial species use the auto-inducers, which supported the “bacterial Esperanto” idea \cite{RN154}, a common language for chemical communication between bacteria. Some caveats have been raised \cite{RN135} and an alternate possibility is raised by our considerations: communicating by diffusible molecules like AI-2 is useful to multicellular communities, and many species have therefore adapted the use of these molecules, often to different ends.
\begin{figure}
\includegraphics{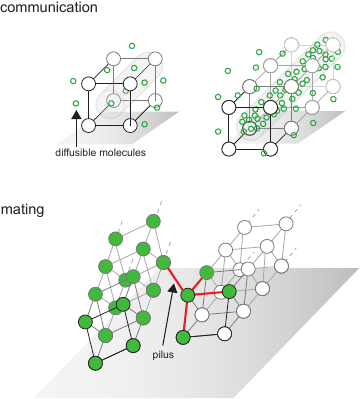}
\caption{Potential behaviors.
\textbf{(Top)} Graphs illustrating the potential role of creating and extending an inner cavity with controlled diameter on cell-cell communication in \textit{E. coli}. Nodes indicate cells and vertices indicate cell-cell connections. Grey cylinder indicates inner cavity and green hexagons indicate diffusible molecules. \textbf{(Bottom)} Graphs illustrating the potential role mating during \textit{E. coli} self-organization. Red vertices indicate the pilus responsible for horizontal DNA transfer and green nodes indicate cells containing transferrable genetic elements.}
\end{figure}

 Of the ways a rod-shaped bacterium might enclose and propagate an internal space, \textit{E. coli}’s may be the most efficient. Rosette formation establishes an inner space in just two cell divisions from the first cell and it orients cells to synchronously grow in tube-like communities based on their natural division geometry. Rosettes on their own do not guarantee inner cavity propagation, which subsequently requires both fimbriae and curli for stability and shape. Considering the developmental geometries of higher organisms, \textit{E. coli} self-organization would unavoidably impact molecular diffusion in a way that could support internal multicellular communication. Though relevant to communication by auto-inducers like AI-2, these points apply to any diffusible molecule excreted into the inner cavity, including metabolic byproducts.

 Secondary metabolites can also communicate between bacteria and play critical roles their responses to different environments \cite{RN155}. For example, indole produced by tryptophan metabolism, acts as a signal in \textit{E. coli} \cite{RN156, RN157} and can induce genes to increase antibiotic tolerance \cite{RN158}. \textit{S. enterica} does not produce indole but also increases its antibiotic tolerance \cite{RN159} in response to physiological concentrations of this molecule \cite{RN160}. Tryptophan catabolism is not activated until more preferred nutrients have been exhausted \cite{RN161, RN162}, suggesting this form of communication is unlikely during multicellular extension, though it could factor during attached dormancy. Questions remain regarding internal communication but the scope seems defined: which molecules are produced and sensed via the inner cavity, how are they interpreted and altered by gene regulation, and what is the dynamic progression of these events throughout community formation? These questions can be investigated by tracking transcriptional dynamics in multicellular communities using fluorescent reporters \cite{RN163} that allow for real-time and longitudinal single-cell measurements. Though outside our current scope, it is worth mentioning that the implications of internal communication for infections may be serious. Internal communication would mean that \textit{E. coli} cells in microscopic communities alter their identities and behaviors (\textit{e.g.}, antibiotic tolerance and colonization) in ways that are completely hidden from classical microbiology methods, including planktonic, colony, and biofilm approaches. While its genetics are well studied, we may know little about \textit{E. coli} as an organism.

\subsection{Mating}
Mating is the process of transferring DNA between organisms. The discovery of genetic exchange between physically-connected \textit{E. coli} transformed biology in 1946 \cite{RN164}. Conjugal paring of individual cells was subsequently observed \cite{RN165}, but the precise role of the pilus remained unclear until just last year when elegant tracking by microscopy verified it served as a conduit for DNA \cite{RN166}. Though often imaged between cell pairs, conjugation occurs in multicellular aggregates as well \cite{RN165}. In fact, the majority of conjugal events may occur in such mating aggregates \cite{RN167}. Formation of multicellular biofilm communities is thought to enhance conjugation and thereby contribute to the spread of antibiotic resistance through horizontal gene transfer \cite{RN168}. However, the precise timing of conjugation, or its dependence on growth or biofilm phase, is not fully clear, and conjugation may be negligible in the mature portions of biofilms \cite{RN169} indicating that it must occur prior to the final stage of biofilm formation. As it encompasses the stages from individual cells to attached dormant communities, \textit{E. coli} self-organization may include a mating phase during which conjugation occurs (\textbf{Fig. 4}, bottom). In clonal communities, conjugation between sister cells harboring identical plasmids would serve little benefit and would be blocked by surface exclusion in cases \cite{RN170, RN168}. Analogous to conjugal pairing of cells \cite{RN165}, but at a larger scale, multicellular chains can come into parallel alignment with each other during extension \cite{RN5}, raising the possibility that \textit{E. coli} may at times mate as clonal multicellular units. Recent innovative approaches \cite{RN171, RN166} might be applied to test this idea. Alternatively, mating could potentially block self-organization as cells presenting adhesive conjugal pili can form multicellular aggregates \cite{RN39, RN38} independently of biofilm/self-organization genes. Even in this case though, conjugation between an individual plasmid-containing donor cell and a plasmid-less recipient cell within a multicellular chain might lead to plasmid dissemination throughout the clonal community (\textbf{Fig. 4}, bottom). Investigating conjugation during \textit{E. coli} self-organization could shed light on the spread of antibiotic resistance genes and the cell-cell interactions of mating.

The “SOS,” DNA-damage response \cite{RN173, RN172} can be induced by conjugation \cite{RN174}. \textit{SulA} is expressed as part of this response and causes filamentation by blocking septation during cell division \cite{RN175, RN176}. Intriguingly, \textit{SulA}-dependent filamentation is observed in the intracellular multicellular communities in some strains \cite{RN177, RN179, RN178}, indicating the host cytoplasm may induce bacterial DNA damage, or alternatively, that genetic exchange may occur in these communities. 

\section{Comparative multicellularity through dynamic graphs}
Not all communities are equal, but organizational equivalences can exist between different organisms. In the past, the emergence of complex multicellularity from unicellular organisms has often been considered a major evolutionary transition \cite{RN11}. Though multicellularity in a simple sense is an inevitable consequence of cell division: unless separated by their own actions or external forces, sister cells would remain together. Cell-cell interactions would be unavoidable, and an organism’s relative multicellularity would reflect the extent to which it used and controlled them. Adding a new behavior might be as easy as altering an existing interaction or using an existing interaction to regulate a gene in a new way. In either case, understanding multicellular evolutionary transitions requires comparing multicellular organization in different communities, which is most effectively begun by considering their dynamic graphs.

For simple communities that have been tracked by microscopy, cell-cell interactions are often easy to see and diagram (even by hand). Cursory comparison suggests the possibility that a defining function of multicellularity is the reliable and ordered propagation of multicellular graphs. If early stages of graph propagation are unreliable or unpredictable, an organism would have little ability (or reason) for sophisticated behaviors at later stages. Relatedly, the reliability of graph propagation is dependent on the type of community organization: some graphs or propagation rules do not lend themselves to reliable dynamic propagation. A cell must know or control its position to properly adapt its behavior and help organize the future cell-cell interactions of a growing community. Such reliability requires controlled progression from one multicellular graph to the next, and self-organization and clonal organization serve clear functions within this context. Robust multicellular dynamics would entail that each multicellular graph represented a meta-stable state, which the system is attracted to but upon reaching is propelled forward to the next state. Only a terminal multicellular graph would be fully stable and would require suspending node propagation (\textit{i.e.}, cell division) and fixing community structure (\textit{e.g.}, by attaching to surfaces or other communities). This in fact is precisely what \textit{E. coli} does by attached dormancy at the end of self-organization. To illustrate some of these points, we discuss examples of \textit{E. coli} multicellular organization by considering their dynamic graphs. The key behaviors enabling robust propagation of multicellular graphs are division, adhesion, repositioning, and volume control, which are complementary and sequentially-ordered during \textit{E. coli} self-organization.

\subsection{Constrained communities}
Using agar, Robert Koch developed the technique of growing bacterial “clones” on solid medium \cite{RN180}. This founded bacteriology and remains how most of us are introduced to the study of bacteria. Physically trapped on a surface, cells divide and have no choice but to grow as multicellular colonies. In this case, cells remain together even if they are actively trying to separate, and their arrangement is determined by their shape and the force of their cellular growth. Division is the key behavior organizing such constrained communities. Regarding their multicellular graphs (\textbf{Fig. 5}), nodes are added by division and cannot be lost; edges are determined by steric effects of surface constraints and cell shape and packing statistics.
\begin{figure*}
\includegraphics{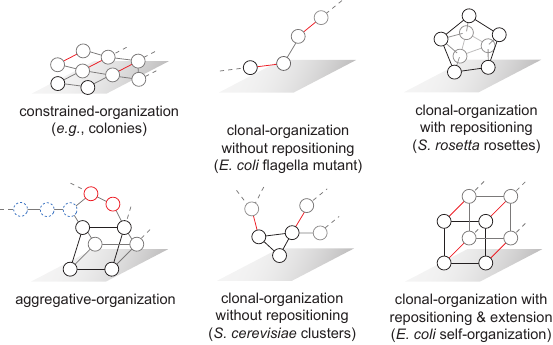}
\caption{\label{fig:wide}Comparative multicellularity.
\textit{Constrained-organization}: community organization is dictated by an external surface, colonies being the canonical example, as cells are forced to grow multicellular-ly. New connections created by cell division are shown in red. Mechanics and packing statistics effectively capture organization in these communities; graph propagation is determined by cell division and physical constraints. Aggregative-organization: community organization is dictated by random collision events and adherence between separate cells or clusters of cells not attached to surfaces. Cells derived from independent clusters are depicted by nodes of differing colors. Thermodynamic equilibrium models treating cells as particles capture key aspects of these communities; graph propagation is random and depends on density and surface properties, which determine the probabilities of collision and adherence, respectively. \textit{Clonal-organization without repositioning}: community organization is determined by the adherence (or lack of separation) of sister cells after division. Without cell repositioning, the angles between cells are not controlled. New connections created by cell division are shown in red for examples: (left) \textit{E. coli} flagella mutant and (right) \textit{S. cerevisiae} clusters. Empirically-identified, organization follows a maximum entropy rule and communities fracture into separate multicellular clusters in a size-dependent manner; graph propagation is determined by cell division and community size, but interactions between nodes is random. \textit{Clonal-organization with repositioning}: community organization is determined by adherence of sister cells and their repositioning relative to one another. Hence, organism-specific multicellular geometries are robustly created by the individual actions of cells, and these communities are considered self-organizing. Community organization is studied through a combination of dynamic observation and genetics; and graph propagation is determined by cell division and repositioning in controlled manner. Rosettes, and similar spherical or circular arrangements of cells, are common motifs in clonal-organization with repositioning and examples are common in developmental biology and multicellular evolution. Their importance is thought to derive from creating an internal cavity that communities can use to coordinate behavior through diffusible molecules. \textit{Clonal-organization with repositioning \& volume control}: community organization is determined by adherence of sister cells and their repositioning relative to each other, followed by directional division and dynamic stabilization of the inner cavity. By controlling the properties of the inner cavity, such communities control how interactions, particular by diffusible molecules, propagate in dynamically changing multicellular communities. As genes that modify cell-cell interactions and the inner cavity can be induced in a stage specific manner, these communities are necessarily studied by multicellular dynamics
; graph propagation occurs in a highly controlled manner and is determined by directionally-coordinated cell division, cell repositioning, volume control, and adaptive gene expression.}
\end{figure*}
Colonies give rise to consistent bulk statistical properties, and during propagation of multicellular graphs it is clear that individual cells have little control their interactions. Steric forces can push cells on top of each other and into three dimensional communities, in which case organization is dependent on adhesins as well as \cite{RN181, RN21}. In higher organisms, cells often adapt their shape and can do so to control cell-cell interactions in constrained communities.

\subsection{Aggregative communities}
The alternative to surface attachment is floating in liquid and having some freedom of motion. In this case, separate cells can come into contact and form aggregates, with a probability determined by their density. Adhesion is the key property organizing aggregative communities. In a strict sense, division is not required and would confound statistical description of aggregation if not controlled for experimentally. Regarding their multicellular graphs (\textbf{Fig. 5}), new nodes and edges between them are added randomly, and propagation of multicellular graphs is mostly left to chance as cells in communities cannot determine either their number or interactions.

In organisms that naturally have clonal organization, high densities introduces aggregation that can mask self-organizing behaviors. Aggregation in these cases can also lead to approximation of aspects of self-organization as seen with adehsin mutants in \textit{E. coli} \cite{RN5} (\textit{Myxococcus} \textit{xanthus} performs a similar aggregate-to-development behavior \cite{RN182}). This idea is also illustrated by organoid models \cite{RN183} and synthetic development in mammals \cite{RN184}, which undergo natural developmental processes if the appropriate aggregates are formed. At low densities, the only way for a cell to form a multicellular community is clonally, by combining division and adhesion, and \textit{E. coli} communities thought to result from aggregation are sometimes shown to result from clonal growth instead \cite{RN40}.

\subsection{Clonal organization without cell repositioning}
Clonal organization combines division and adherence and is distinct from the clonality of colonies, which are forced to grow as communities due to external constraints. There are many microbial examples, in bacteria and yeast, where cells remain adhered after division in liquid media but do not actively reposition  \cite{RN124, RN12, RN185, RN73}. Flagellar mutants of \textit{E. coli} illustrate this type of organization, cells remain adhered at their division site and the angle between cells is unregulated. Considering their multicellular graphs (\textbf{Fig. 5}), new nodes are added by division and can be lost by reducing adhesion. Edges are determined by division and random fluctuations that alter the angle between cells. Clonal organization provides key advantages for reliably propagating multicellular graphs as such communities can control their numbers, which is a challenge for constrained and aggregative communities. Without cell repositioning however, interactions between cells will often be random and follows a maximum-entropy rule \cite{RN186}, meaning they are maximally disorder within the physical constraints posed by cell shape and size. Hence, the interactions of such communities are not controlled and cannot be reliably propagated from one graph to the next. Organisms that lack a repositioning mechanism will have difficulty reliably propagating their multicellular graphs and their ability to perform sophisticated multicellular behaviors will likely be limited.

\subsection{Clonal organization with cell repositioning}
These communities add cell propulsion or motility to clonal organization and are common in development. Regarding their multicellular graphs (\textbf{Fig. 5}), new nodes are added by division; and edges representing physical interactions are determined the repositioning of cells. Such behaviors can create multicellular graphs where the number, positions, and physical interactions of cells are reliably determined. Often, such communities are spherical or circular and are commonly described as rosettes, for example in \textit{S. rosetta} \cite{RN90} or during \textit{E. coli} self-organization. In cases, cells will sense when they reach such initial geometries and adapt their behavior, thereby determining new rules for the next propagation stage of their multicellular graphs. These three behaviors: division, adhesion, and cell movement appear necessary, and nearly sufficient, to creating multicellular graphs that can be reliably propagated.

But cells in communities also unavoidably interact by diffusible molecule. By enclosing an internal space, rosettes can enable multicellular communication by such molecules. An inner space would technically occur in any multicellular community, but rosettes typically position each cell to have equivalent access to the inner space and to the external environment, usually necessary for nutrients and growth. The next step of graph propagation though is critical in determining if such diffusional interactions are maintained and propagated or diminished. As discussed above (internal communication), spherical growth quickly diminishes interaction by diffusible molecules, and with it the potential for multicellular communication. To enable interactions by diffusible molecules, and propagate these interactions from graph to graph in the future, requires controlling the properties of an inner volume. Therefore, reliable propagation of multicellular graphs also requires regulated expansion of the inner volume. As mentioned above, the inner volume’s diameter is a critical property for diffusion and extending as tubes is a multicellular solution to the problem of chemical diffusion.

\subsection{Clonal organization with re-positioning and volume control}
This type of organization combines division, adhesion, repositioning, and inner volume control, which together may represent a foundational set of behaviors for reliable propagation of multicellular graphs. Variations on these behaviors could produce a great variety of graph progressions and hence types of development. For example, branching morphogenesis would lead to propagation of alternative inner volumes, and cellular differentiation or specialization would result from cells precisely knowing their cell-cell interactions. Without these behaviors, the ability to regulate graph progression or complex multicellularity is hard to imagine. Of course, as organisms increase in scale, larger-scale forms of communication like signaling by hormones (endocrine system) and neurons (nervous systems) become essential.

The simplest means of volume control for regulating interactions by diffusible molecules is to extend communities with approximately constant-diameter inner volumes. This habit is seen throughout developmental biology and during \textit{E. coli} self-organization. Post-rosette formation, \textit{E. coli} produces extracellular matrix components, fimbriae and curli, to propagate the inner space between cells and regulate its shape. The extracellular matrix in \textit{E. coli} may effect diffusion through this space, similarly to its role during lumenogenesis in higher organisms. In \textit{E. coli} self-organization, it is unclear how much internal communication takes place during extension, though the 8 cell generations transpiring as communities grow from 4 to $\sim$1,000 cells would seem ample time. In any case, controlling the inner space between cells provides a way to control the cell-cell interactions by diffusible molecules and therefore to reliable propagate cell-cell interactions mediate by such molecules.

Regarding multicellular graphs (\textbf{Fig. 5}), nodes are added by division; initial edges are added through a combination of adhesion and repositioning, and are subsequently added and altered dynamically due to the adaptation of cells to each successive multicellular graph. Hence, such communities are potentially capable of reliable propagation from one multicellular graph to the next, until they reach their developmental endpoint, or terminal graph. The diversity of biological molecules is rich and communicating by their diffusion is powerful. Studying how a volume’s properties control diffusion is a fundamental part of chemical engineering curricula (\textit{e.g.}, Transport and Separations), though for practical reasons, these volumes are \textit{static} and not changing. It is impressive that \textit{E. coli} and other organisms appear to use similar principles to regulate \textit{dynamic} volumes and thereby elegantly enable multicellular communication and long-term self-organization with it. Our trusty companion \textit{E. coli} had lessons in control theory for us after all \cite{RN187}.

Evidently and intuitively, the multicellular behaviors of division, adhesion, repositioning, and volume control appear to be well-ordered. Generally, an organism cannot perform any of them without performing the preceding ones first, and more sophisticated behaviors like internal communication and differentiation may be shown in the future to depend on them. From the perspective of comparative multicellularity, the verified behaviors of \textit{E. coli} self-organization would indicate it has greater relative multicellularity than many examples of constrained-organization or aggregative-organization, and even than some simple multicellular eukaryotes. Though it appears to perform these behaviors effortlessly, \textit{E. coli} makes its home in a very successful multicellular organism and may not benefit from further complexity. Using dynamic graphs to contextualize the behaviors of other organisms will likely add to and refine the ideas presented here and has potential to uncover principles both fundamental and specific to the theory of multicellularity.
\vspace{5 mm}
\section{Outlook}
\textit{E. coli} self-organization includes a unique set of multicellular behaviors with similarities to higher organisms. Investigation is ongoing and studying \textit{E. coli’s} multicellular dynamics (cell-cell interactions and cell responses to them) will uncover additional behaviors in the future. Interpreting and comparing multicellular dynamics (of \textit{E. coli} and other organisms) through dynamic graphs will reveal fundamental principles of multicellularity, which will be essential for understanding a variety of diseases and engineering new multicellular possibilities.
\bibliography{SelfOrgan}
\end{document}